\documentclass[11pt,a4paper]{article}
\usepackage[latin1]{inputenc}
\usepackage{amsmath}
\usepackage{amsfonts}
\usepackage{amssymb}

\usepackage{geometry}
\usepackage[tt=false]{libertine}
\usepackage[T1]{fontenc}

\usepackage{subcaption}

\title{Reproducibility Report for the Paper:\\\emph{QN-based Modeling and Analysis of Software Performance Antipatterns for Cyber-Physical Systems}}
\author{Alessandro Pellegrini\\a.pellegrini@ing.uniroma2.it}
\date{February 3, 2021}

\usepackage{tikz,pgfplots,pgfplotstable}
\usepackage{amsmath}
\usepackage{graphicx}
\usepackage{float}

\usepackage{xcolor}
\usepackage{mdframed}
\usepackage[labelfont=bf]{caption}

\usepackage[hidelinks]{hyperref}

\begin{document}
	
\maketitle

\abstract{
The authors have uploaded their artifact to Zenodo, which ensures a long-term retention of the artifact. The artifact allows to re-run the experiments very smoothly, and the dependencies are well documented. The process to regenerate data for the figures and tables in the paper completes, and all results are reproducible.

This paper can thus receive the \textit{Artifacts Available} badge. The software in the artifact runs correctly with no trouble, and is relevant to the paper, thus deserving the  \textit{Artifacts Evaluated---Functional} badge. Given the successful reproduction of all figures and tables, the \textit{Results Reproduced} badge can be assigned.}

\section{Introduction}

The paper \emph{QN-based Modeling and Analysis of Software Performance Antipatterns for Cyber-Physical Systems}~\cite{paper} by Riccardo Pinciroli, Connie U. Smith, and Catia Trubiani presents a methodology to understand software performance problems in cyber-physical systems.

The authors discuss several \textit{software performance antipatters}, namely bad practice when designing and implementing software systems, that can severely affect the performance. Queuing network-based performance models are proposed, to illustrate what could be the effects on performance of these antipatterns.

The authors perform a large simulation-based analysis of their models, to demonstrate their usefulness at spotting these problems early in a software system design.

\section{Replication of Computational Results}

In this section, I will comment the process associated with the reproducibility of this artifact, highlighting the reasons behind the decisions which have already been highlighted in the abstract. Original figures in the paper are published with permission from the authors.

\subsection{Software download and installation}

The authors have provided a link to a permanent repository on Zenodo (\url{https://zenodo.org/record/4495665}), making it permanently available (\textbf{Artifacts Available badge}).

The amount of dependencies required to run the artifact is extremely reduced. Simulations are driven by a set of Python 3 scripts, which leverage the Java Modelling Tools~\cite{JMT} version 1.1.0. Python dependencies are mainstream: numpy, pandas, and matplotlib. To re-generate the results, there is a subsequent post-processing phase, which is based on a couple of Jupyter notebooks.

\subsection{Quality of the artifact}

The artifact is relevant to the associated paper, as it allows to re-generate all the tables and all the figures, it is accompanied by a clear README file which illustrates how to smoothly regenerate the results, and it is easy to use. 
There is a part of the results, related to two columns in Table 6 in the paper, which have been obtained using a commercial tool named SPE$\cdot$ED~\cite{SPEED}. Given that this is a commercial closed-source tool, it is not included in the artifact. Anyhow, also these results have been reproduced, thanks to the authors giving the possibility to engage in a live videoconferencing session to exercise the SPE$\cdot$ED tool (\textbf{Artifacts Evaluated---Functional}). 

The authors have also provided the results of their simulation runs in the artifact. I can confirm that by relying on this data, the post-processing and plotting modules do work, and the figures are 1:1 with respect to those in the paper. Nevertheless, this reviewer has deleted all the results in the artifact, to exercise the whole software package.

\subsection{Reproducing the experiments}

Reproduction of the results has been carried out on a Linux machine, using the Arch Linux distribution, running kernel 5.10.4. The used interpreter was python 3.9.1. I have used JMT version 1.0.4---a lower version than the one used by the authors, but it caused no compatibility issues. Additional software versions used for the reproducibility are as follows. Jupyter core 4.6.3, jupyter notebook 6.1.6, ipython 7.19.0, ipykernel 5.4.2, jupyter client 6.1.7.

The artifact is easy to run, and all scripts required to support the reproducibility are provided (\textbf{Artifacts Evaluated---Functional}). To run all the simulations, it has taken around 6 hours on a quad-core Intel i7-7600U CPU.

In this section, I provide a short description of my reproducibility results. There is a total of 13 Figures and 6 Tables in the original paper.
\textbf{Tab. 1}, \textbf{Tab. 2}, \textbf{Tab. 3}, \textbf{Tab. 4}, and \textbf{Tab. 5} in the paper do not present research results, and therefore must not be reproduced.
\textbf{Fig. 1}, \textbf{Fig. 2}, \textbf{Fig. 4}, \textbf{Fig. 5}, \textbf{Fig. 6}, \textbf{Fig. 8}, \textbf{Fig. 9}, and \textbf{Fig. 10} in the paper do not present results and need not to be reproduced.

In Figure~\ref{fig3} we report the reproduced results, compared to the original figure in the paper. The simulation has been run with a different random seed, nevertheless the obtained outcome is comparable both in magnitude and in variance.

The reproduced results related to Figure~\ref{fig7a} and Figure~\ref{fig7b} are similarly comparable. In particular, it is interesting to observe the similar trend for the variance, and the sweet spot around 0.02. Similarly, the results for Figure~\ref{fig7c} observe a similar trend and a similar variance.

Also the trends for Figures~\ref{fig11}, \ref{fig12a}, \ref{fig12b} are comparable. With respect to Figure~\ref{fig12c}, the reproduced results look better than the ones in the paper: they are stabler and show a more reduced variance, which is an indication of the soundness of the proposal in the original paper. The same applies for Figures~\ref{fig12d} and \ref{fig13a}.

With respect to Figure~\ref{fig13b}, the reproduced results show a similar trend, but at the same time they appear to be less noisy and stabler. This is again an indication of the robustness of the proposal in the original paper.

In Table~\ref{thetab} we report results partially obtained via the artifact. As mentioned before, the columns related to EG in the table are related to proprietary software not included in the artifact. The results have been reproduced in a live videoconferencing session with the authors. Care has been taken in configuring the model with the same input parameters discussed in the paper. 
As it can be observed in the reproduced table, the overall results are comparable, while some of the results look even better than in the original paper.

\renewcommand{\thefigure}{3}
\begin{figure}[ht]
	\centering
	\begin{subfigure}{0.45\textwidth}
		\centering
		\includegraphics[width=\textwidth]{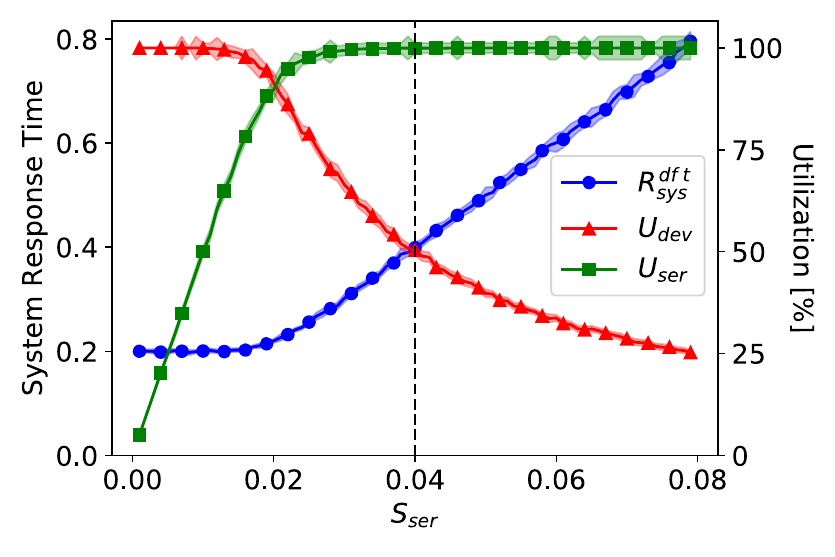}
		\caption{Original.}
	\end{subfigure}
	\hfill
	\begin{subfigure}{0.45\textwidth}
		\centering
	    \includegraphics[width=\textwidth]{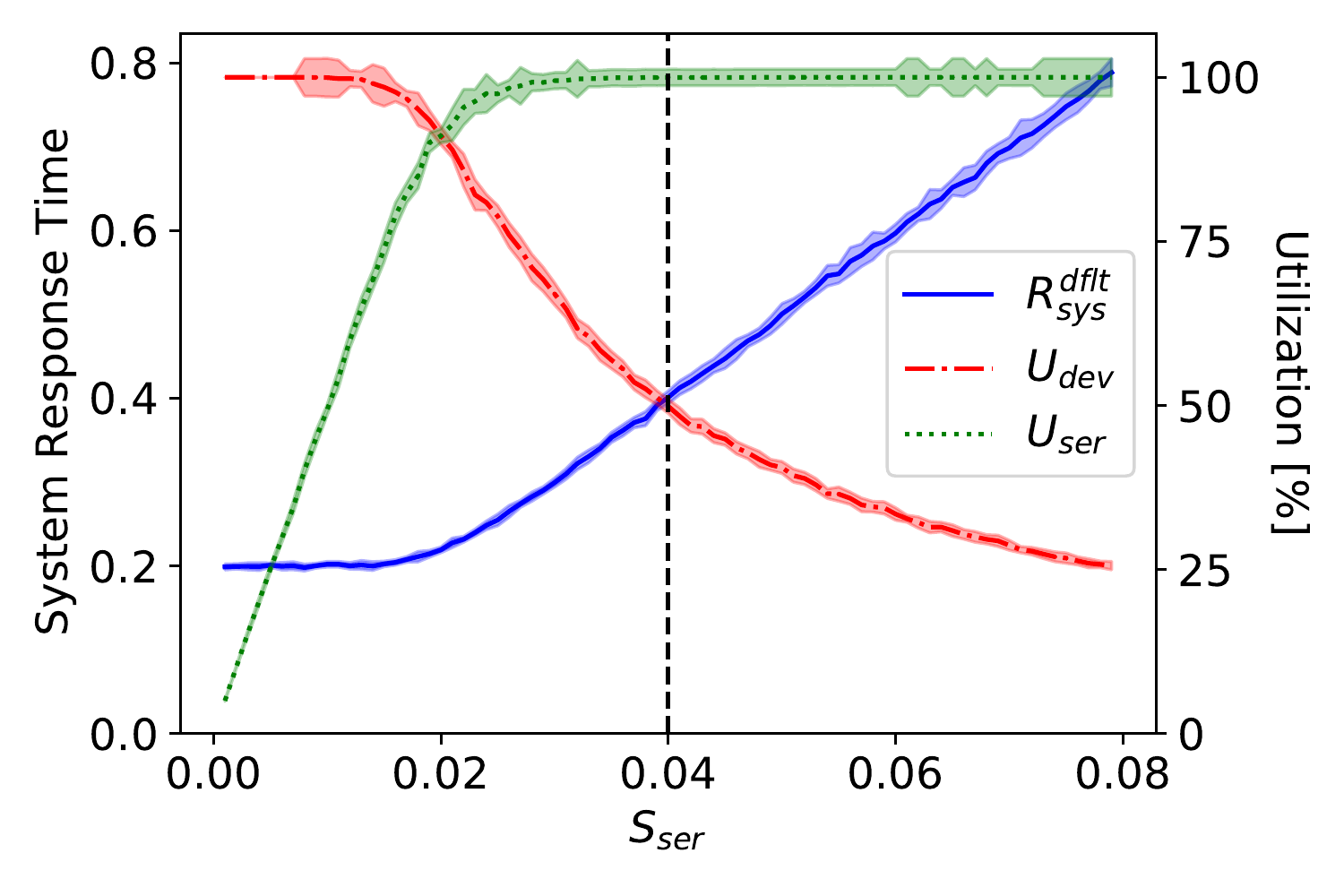}
		\caption{Reproduced.}
	\end{subfigure}
	\caption{Model-based performance analysis results of the
baseline queuing network.}
	\label{fig3}
\end{figure}

\renewcommand{\thefigure}{7a}
\begin{figure}[ht]
	\centering
	\begin{subfigure}{0.45\textwidth}
		\centering
		\includegraphics[width=\textwidth]{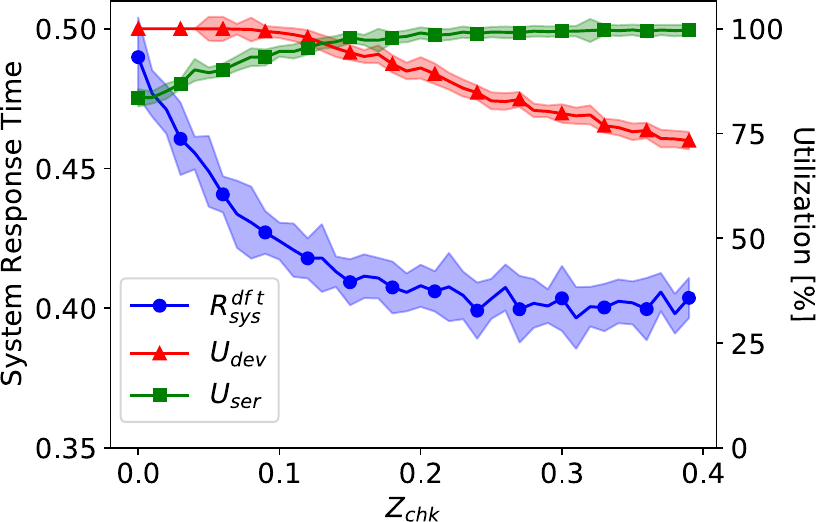}
		\caption{Original.}
	\end{subfigure}
	\hfill
	\begin{subfigure}{0.45\textwidth}
		\centering
	    \includegraphics[width=\textwidth]{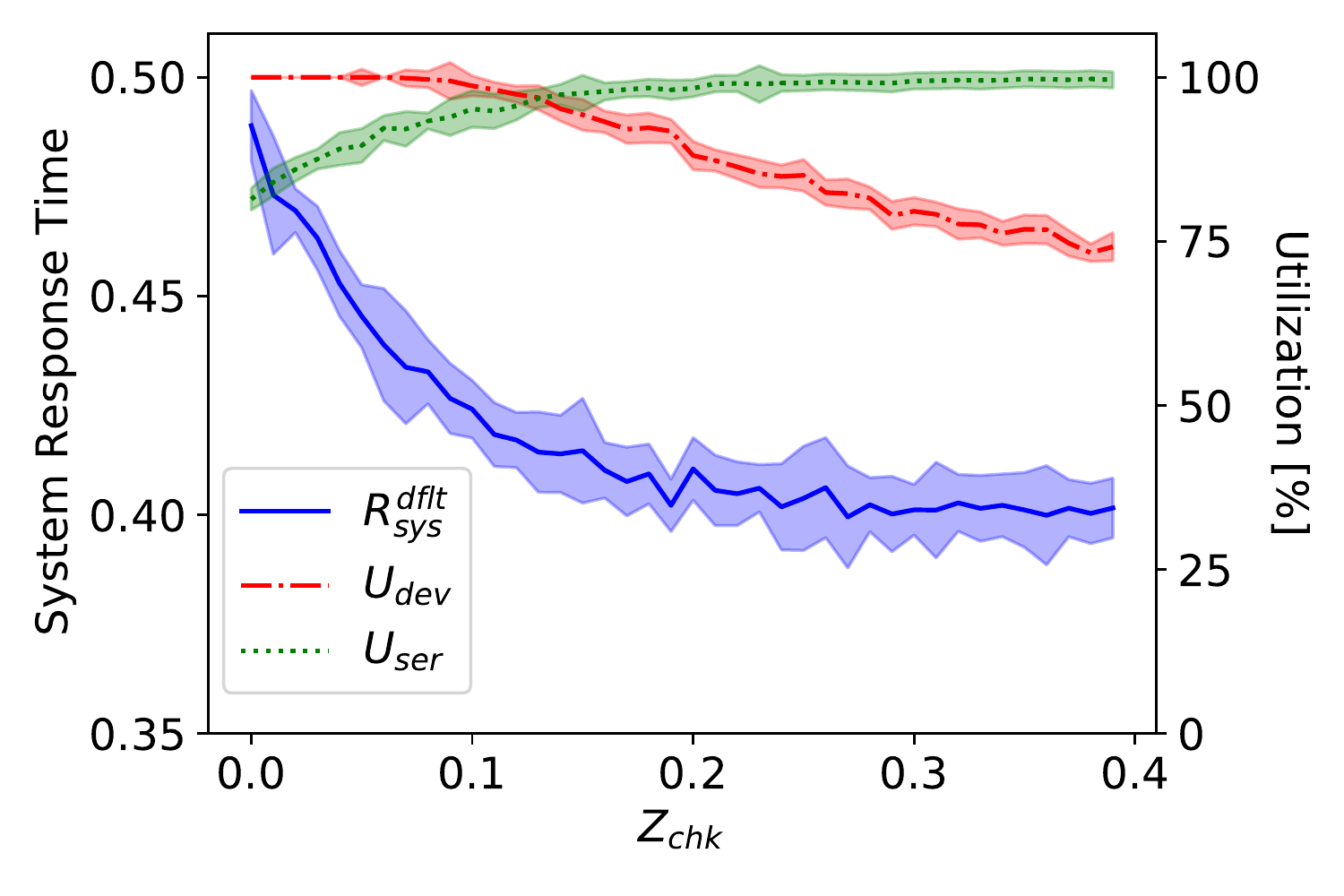}
		\caption{Reproduced.}
	\end{subfigure}
	\caption{Effect of the three performance antipatterns on the baseline system: Are we there yet?}
	\label{fig7a}
\end{figure}

\renewcommand{\thefigure}{7b}
\begin{figure}[ht]
	\centering
	\begin{subfigure}{0.45\textwidth}
		\centering
		\includegraphics[width=\textwidth]{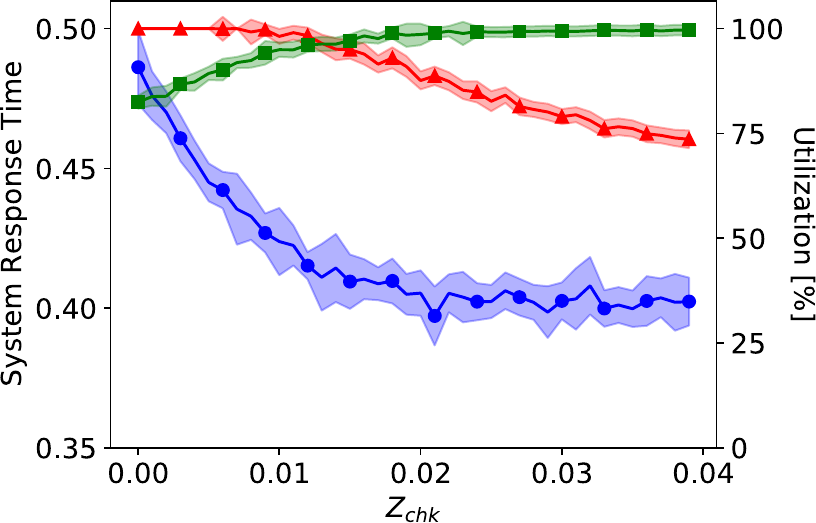}
		\caption{Original.}
	\end{subfigure}
	\hfill
	\begin{subfigure}{0.45\textwidth}
		\centering
	    \includegraphics[width=\textwidth]{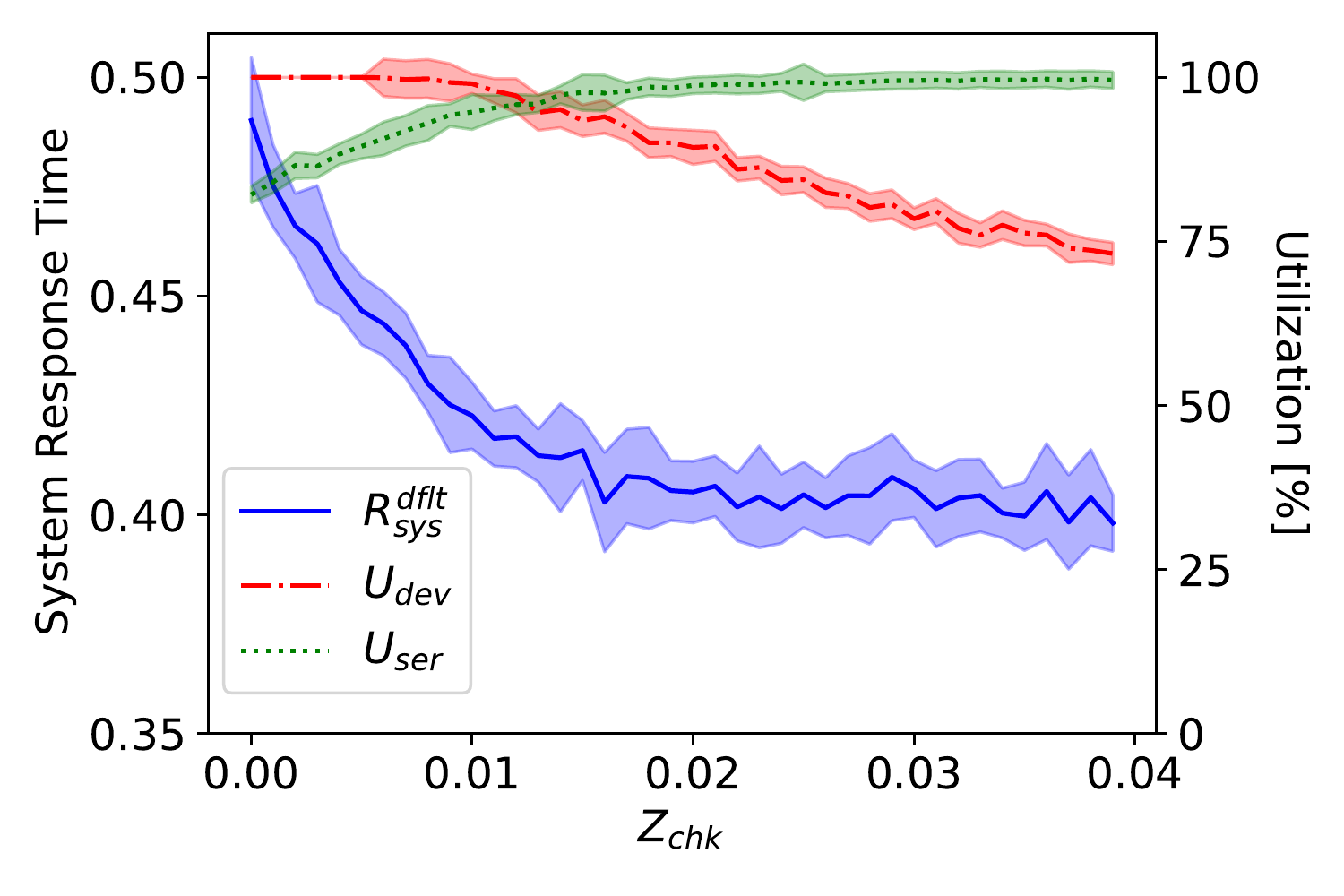}
		\caption{Reproduced.}
	\end{subfigure}
	\caption{Effect of the three performance antipatterns on the baseline system: Is Everything OK?}
	\label{fig7b}
\end{figure}

\renewcommand{\thefigure}{7c}
\begin{figure}[ht]
	\centering
	\begin{subfigure}{0.45\textwidth}
		\centering
		\includegraphics[width=\textwidth]{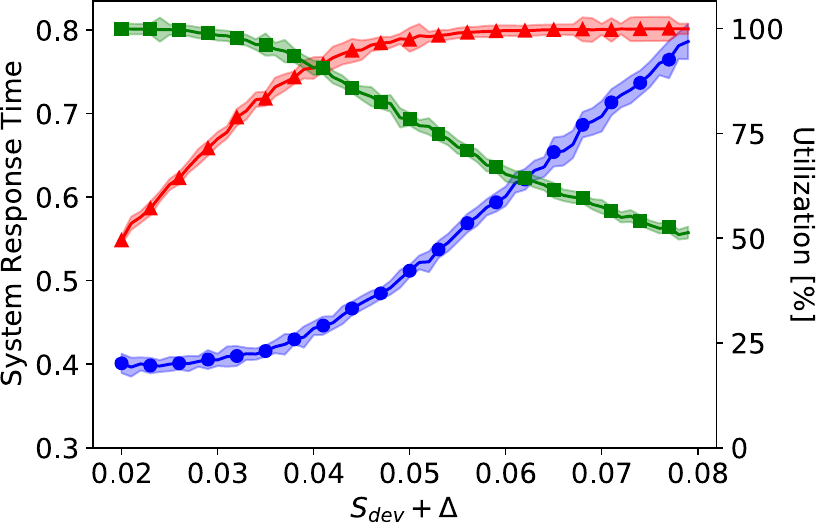}
		\caption{Original.}
	\end{subfigure}
	\hfill
	\begin{subfigure}{0.45\textwidth}
		\centering
        \includegraphics[width=\textwidth]{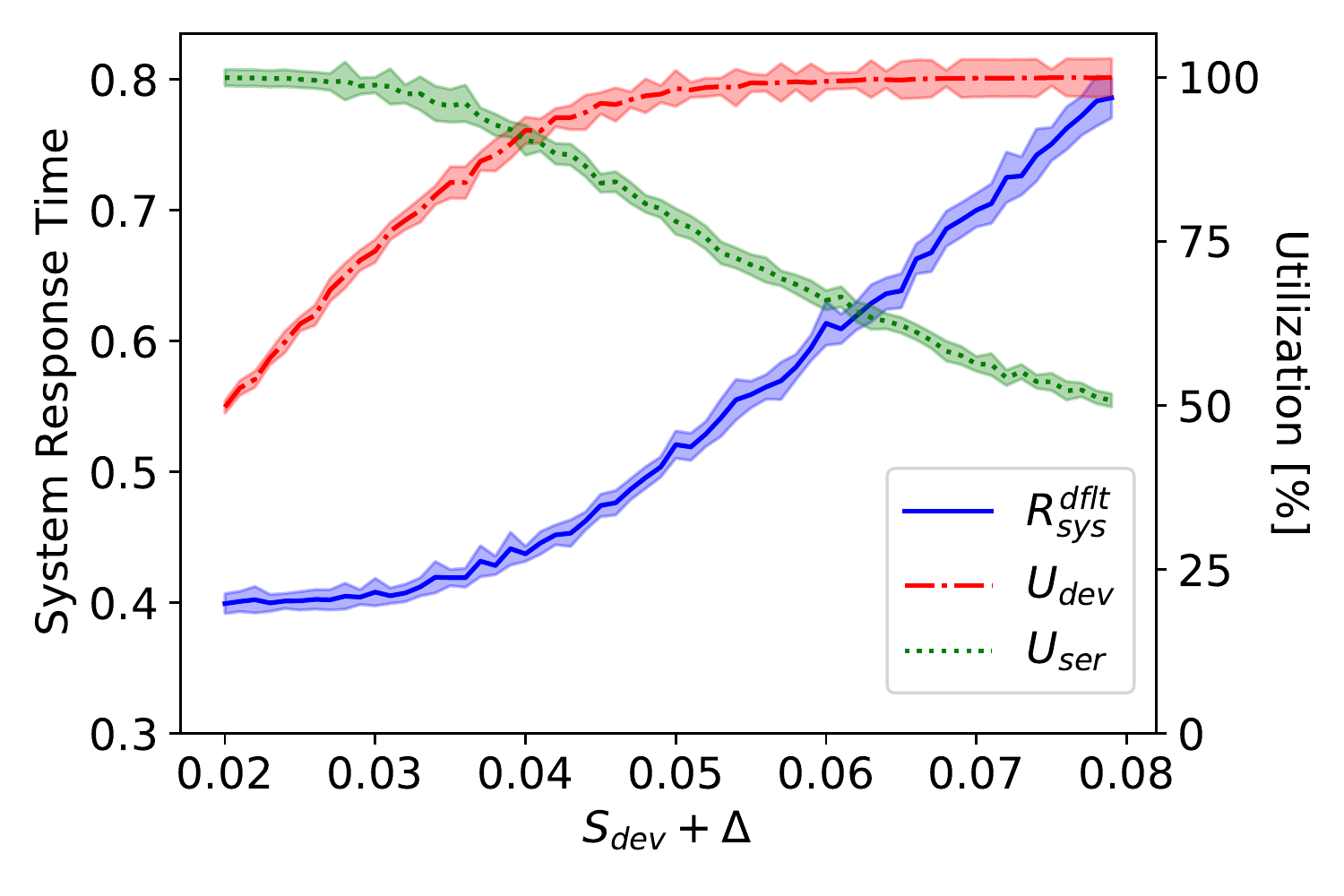}
		\caption{Reproduced.}
	\end{subfigure}
	\caption{Effect of the three performance antipatterns on the baseline system: Where Was I?}
	\label{fig7c}
\end{figure}

\renewcommand{\thefigure}{11}
\begin{figure}[ht]
	\centering
	\begin{subfigure}{0.45\textwidth}
		\centering
		\includegraphics[width=\textwidth]{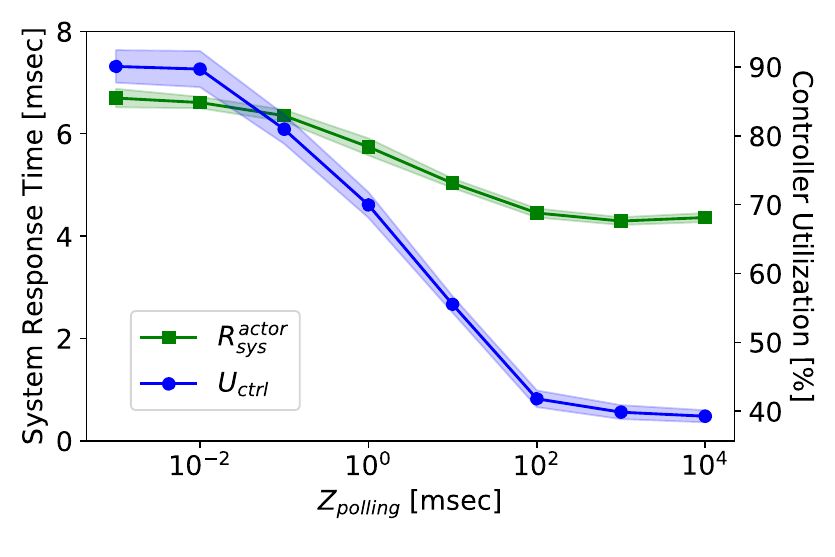}
		\caption{Original.}
	\end{subfigure}
	\hfill
	\begin{subfigure}{0.45\textwidth}
		\centering
	    \includegraphics[width=\textwidth]{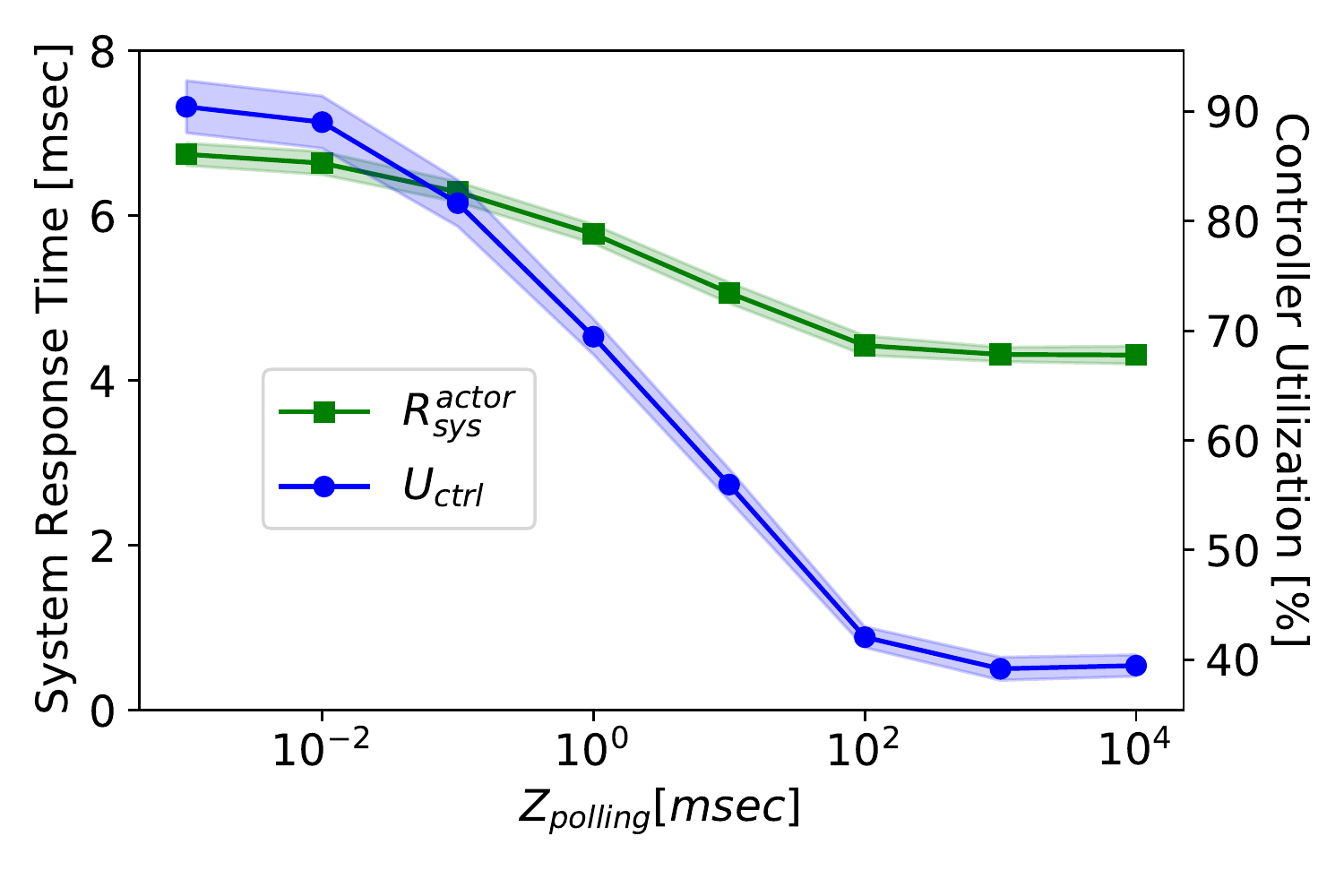}
		\caption{Reproduced.}
	\end{subfigure}
	\caption{Effect of \textit{Are We There Yet?} software performance antipattern on the sensor net system.}
	\label{fig11}
\end{figure}

\renewcommand{\thefigure}{12a}
\begin{figure}[ht]
	\centering
	\begin{subfigure}{0.45\textwidth}
		\centering
		\includegraphics[width=\textwidth]{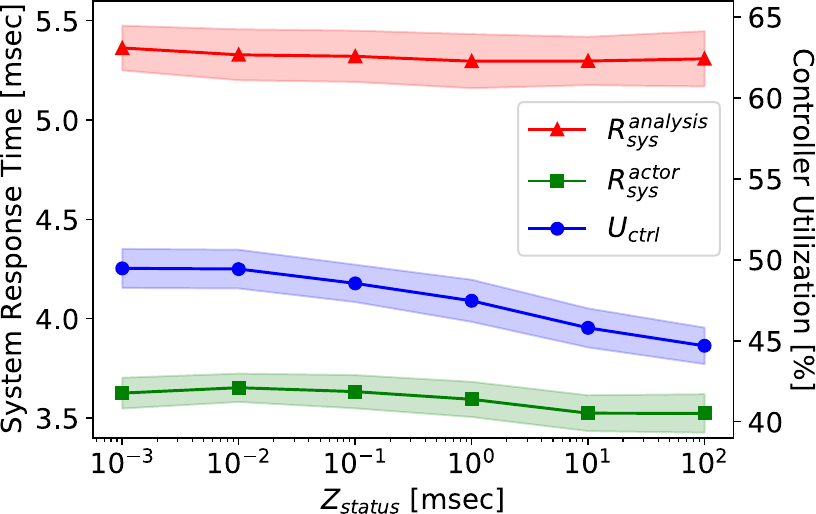}
		\caption{Original.}
	\end{subfigure}
	\hfill
	\begin{subfigure}{0.45\textwidth}
		\centering
	    \includegraphics[width=\textwidth]{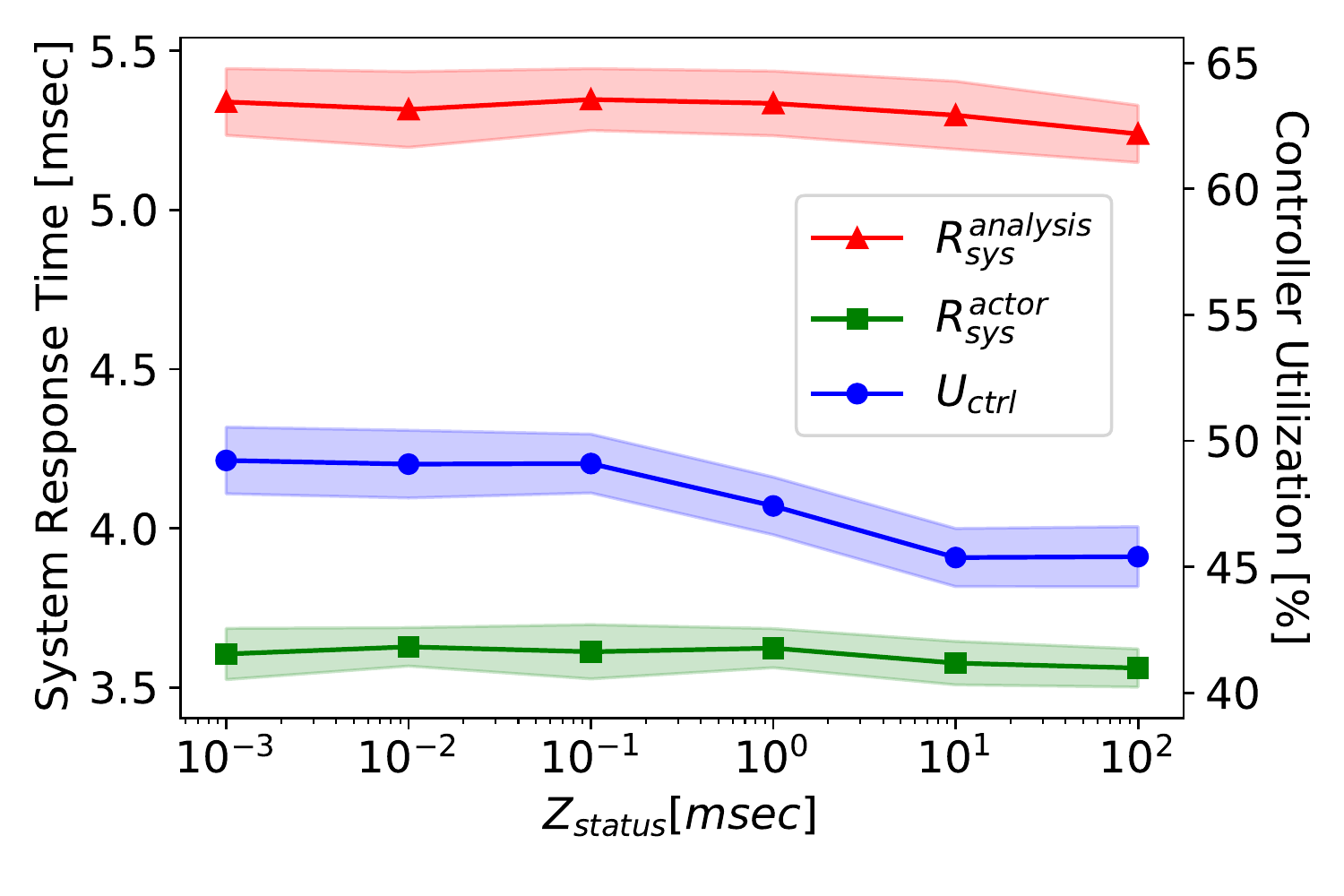}
		\caption{Reproduced.}
	\end{subfigure}
	\caption{Effect of \textit{Is Everything OK?} software performance antipattern on the sensor net system when checked devices (i.e., sensors) do not return exceptions. The performance of the \textit{Controller} is evaluated for different numbers of \textit{Status jobs}. $N_{status} = 1$.}
	\label{fig12a}
\end{figure}

\renewcommand{\thefigure}{12b}
\begin{figure}[ht]
	\centering
	\begin{subfigure}{0.45\textwidth}
		\centering
		\includegraphics[width=\textwidth]{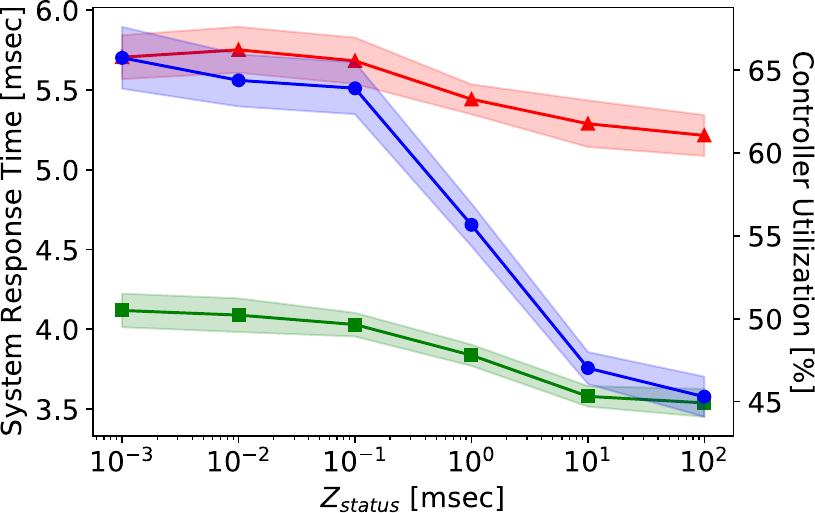}
		\caption{Original.}
	\end{subfigure}
	\hfill
	\begin{subfigure}{0.45\textwidth}
		\centering
	    \includegraphics[width=\textwidth]{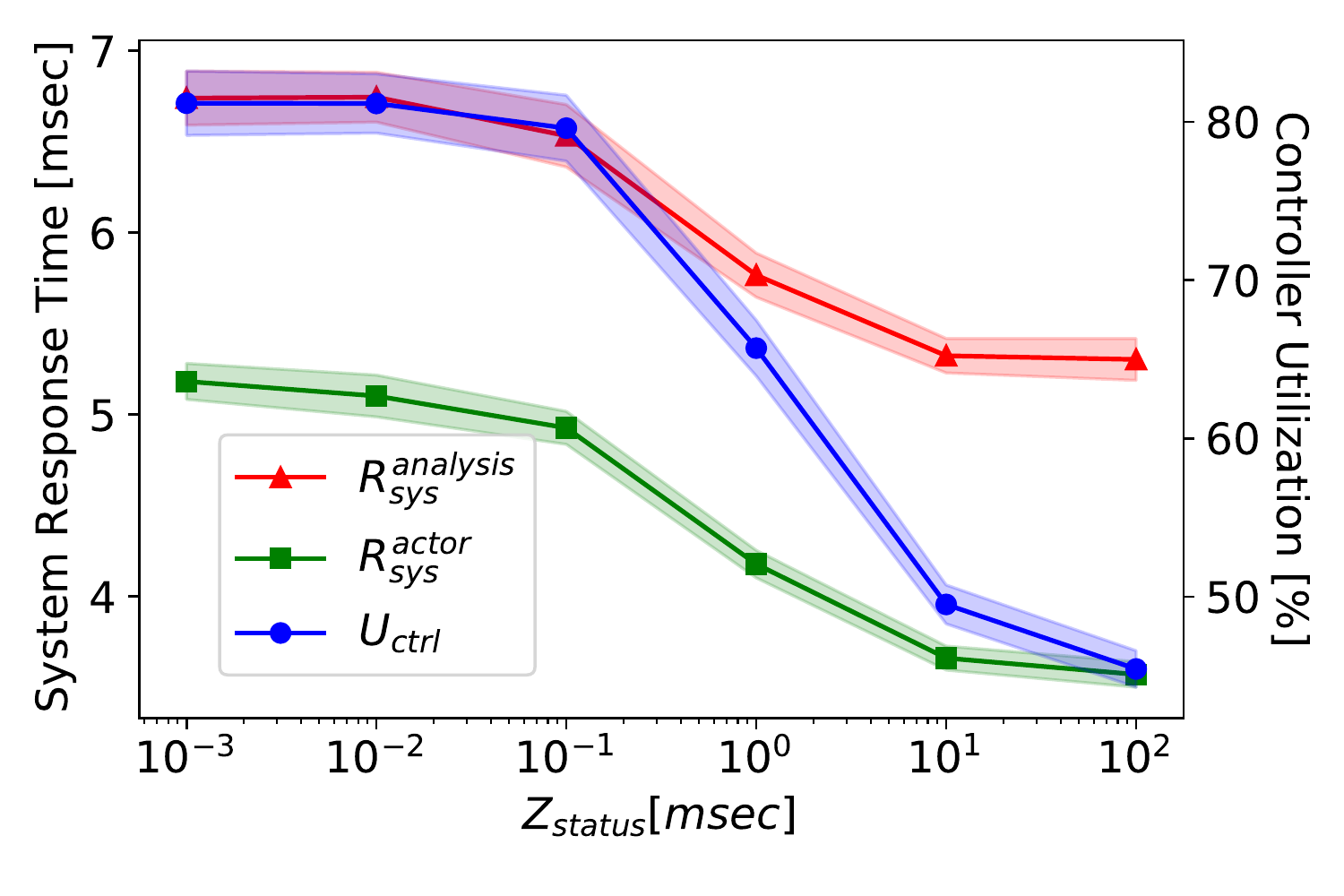}
		\caption{Reproduced.}
	\end{subfigure}
	\caption{Effect of \textit{Is Everything OK?} software performance antipattern on the sensor net system when checked devices (i.e., sensors) do not return exceptions. The performance of the \textit{Controller} is evaluated for different numbers of \textit{Status jobs}. $N_{status} = 5$.}
	\label{fig12b}
\end{figure}

\renewcommand{\thefigure}{12c}
\begin{figure}[ht]
	\centering
	\begin{subfigure}{0.45\textwidth}
		\centering
		\includegraphics[width=\textwidth]{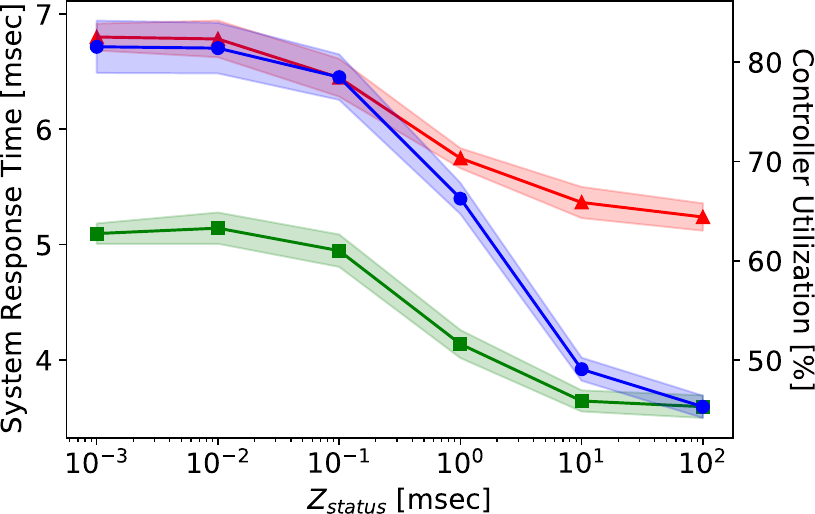}
		\caption{Original.}
	\end{subfigure}
	\hfill
	\begin{subfigure}{0.45\textwidth}
		\centering
	    \includegraphics[width=\textwidth]{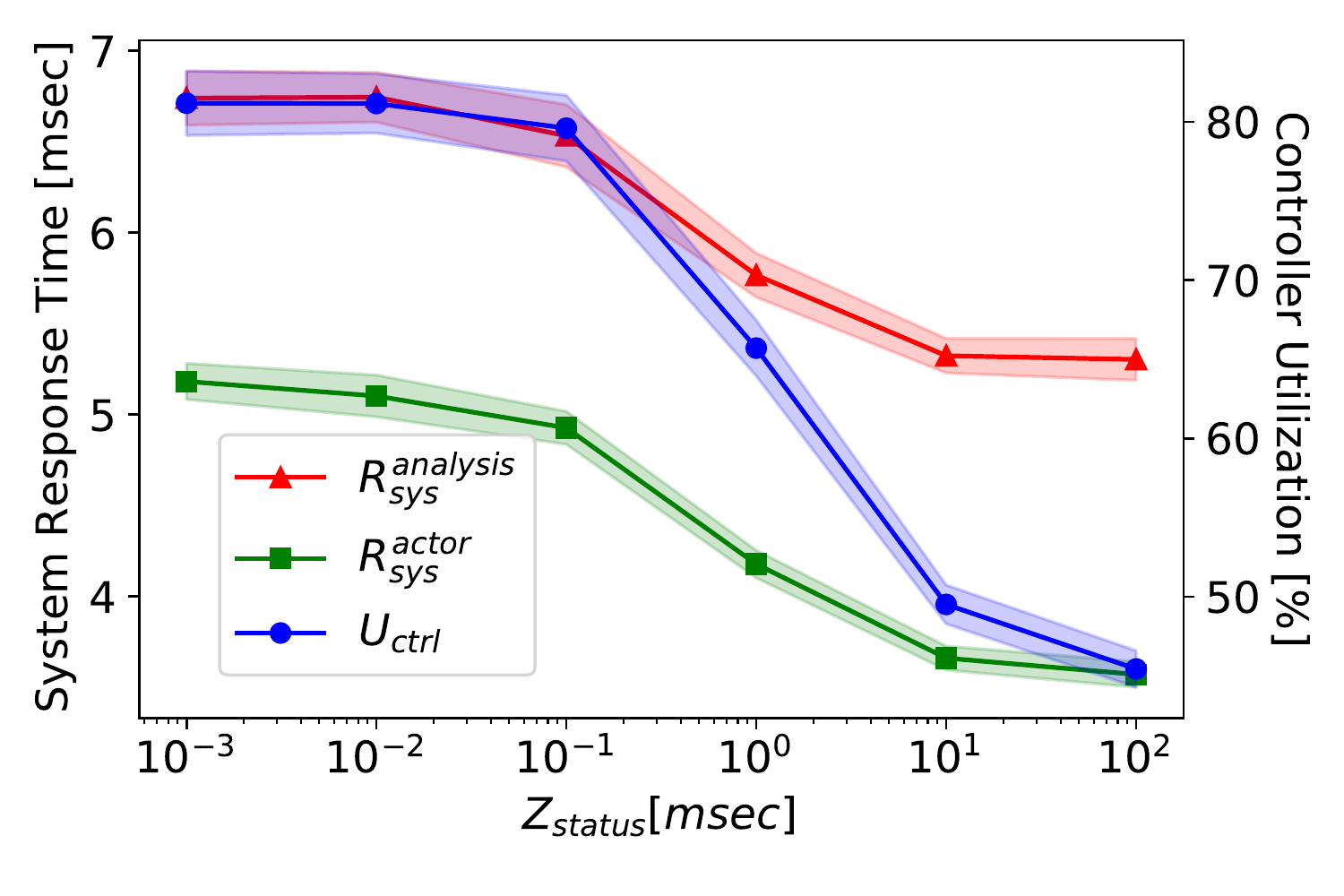}
		\caption{Reproduced.}
	\end{subfigure}
	\caption{Effect of \textit{Is Everything OK?} software performance antipattern on the sensor net system when checked devices (i.e., sensors) do not return exceptions. The performance of the \textit{Controller} is evaluated for different numbers of \textit{Status jobs}. $N_{status} = 10$.}
	\label{fig12c}
\end{figure}

\renewcommand{\thefigure}{12d}
\begin{figure}[ht]
	\centering
	\begin{subfigure}{0.45\textwidth}
		\centering
		\includegraphics[width=\textwidth]{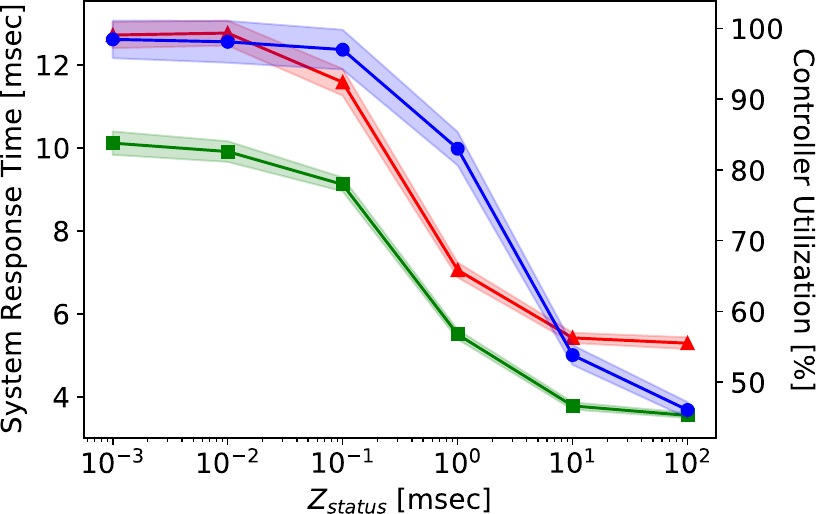}
		\caption{Original.}
	\end{subfigure}
	\hfill
	\begin{subfigure}{0.45\textwidth}
		\centering
	    \includegraphics[width=\textwidth]{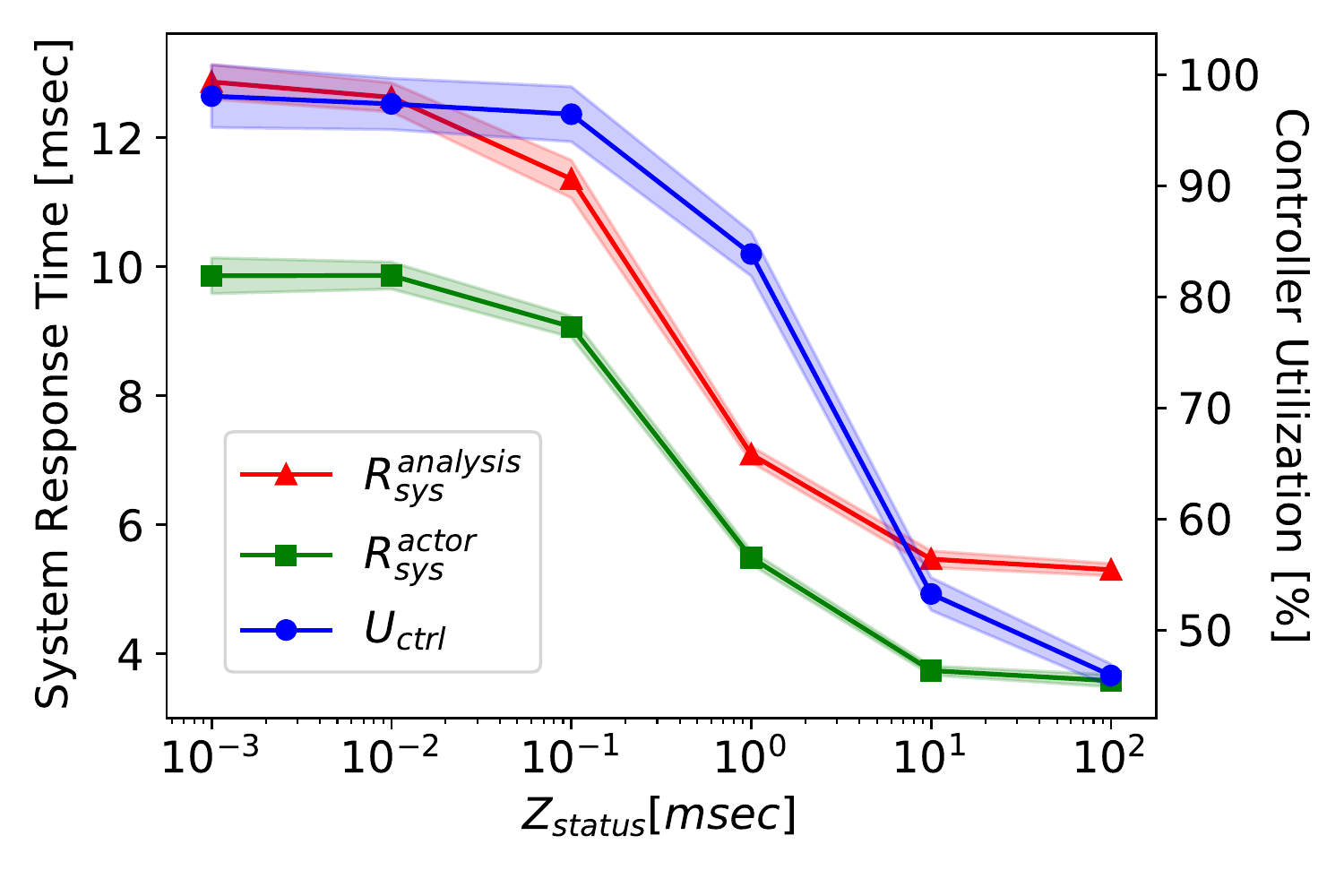}
		\caption{Reproduced.}
	\end{subfigure}
	\caption{Effect of \textit{Is Everything OK?} software performance antipattern on the sensor net system when checked devices (i.e., sensors) do not return exceptions. The performance of the \textit{Controller} is evaluated for different numbers of \textit{Status jobs}. $N_{status} = 20$.}
	\label{fig12d}
\end{figure}

\renewcommand{\thefigure}{13a}
\begin{figure}[ht]
	\centering
	\begin{subfigure}{0.45\textwidth}
		\centering
		\includegraphics[width=\textwidth]{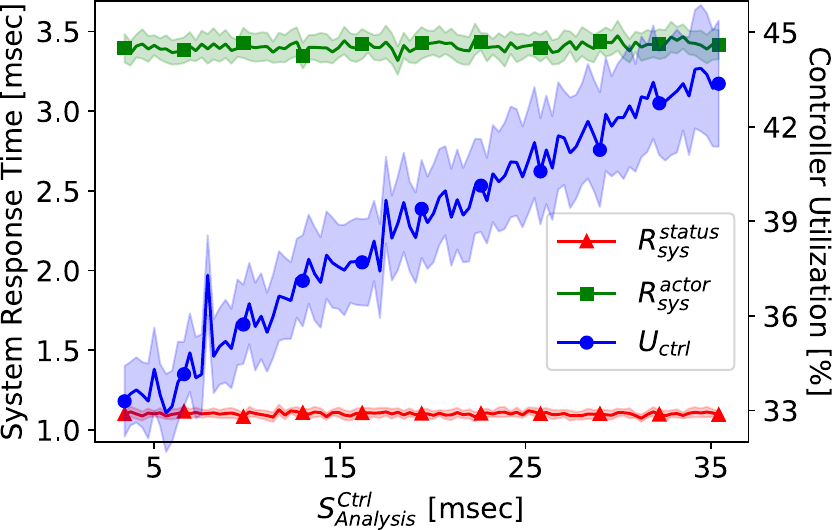}
		\caption{Original.}
	\end{subfigure}
	\hfill
	\begin{subfigure}{0.45\textwidth}
		\centering
	    \includegraphics[width=\textwidth]{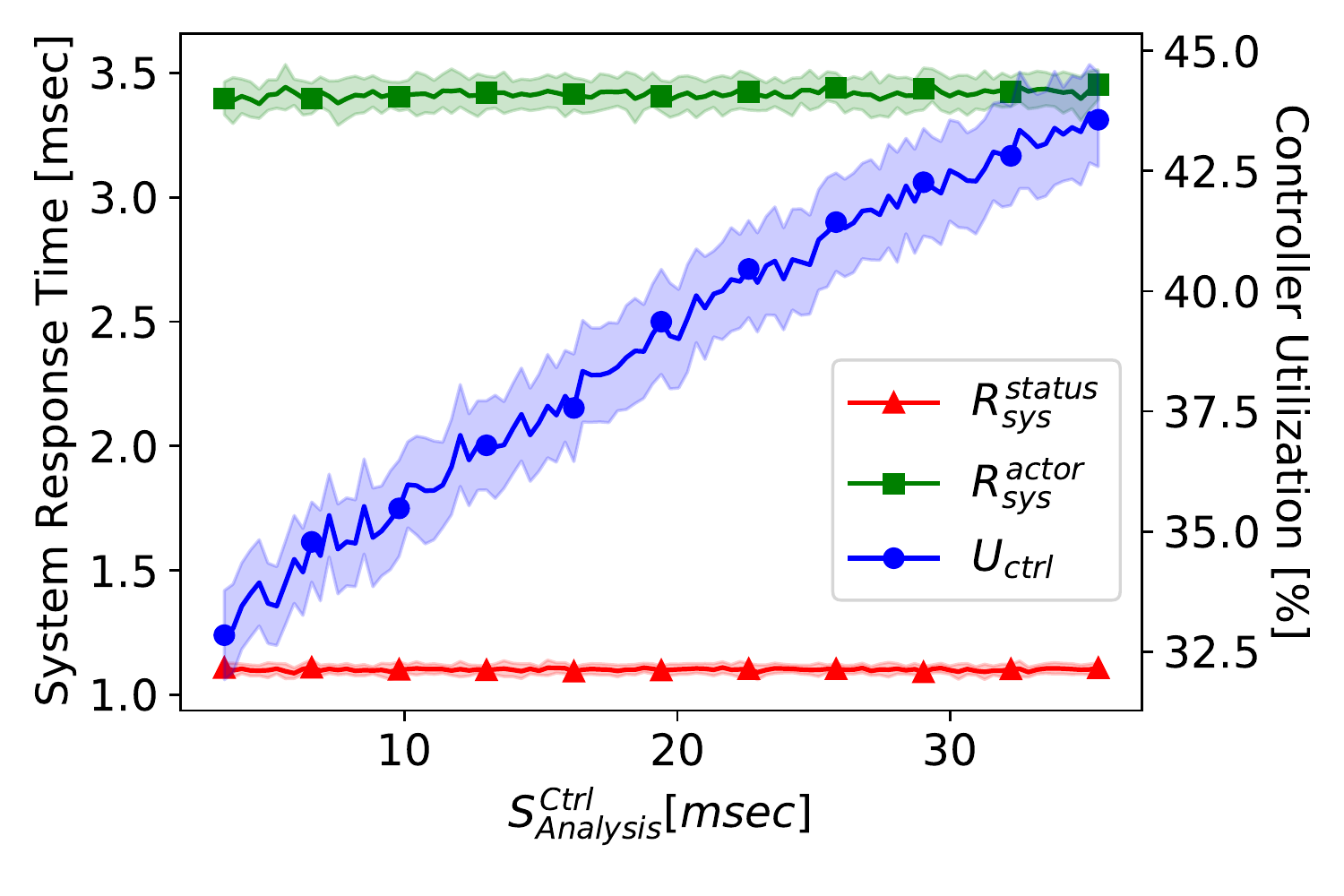}
		\caption{Reproduced.}
	\end{subfigure}
	\caption{Effect of \textit{Where Was I?} software performance antipattern on the sensor net system. Results are obtained considering only 1 Actor and 1 Sensor in the system. Other input parameters are the same as those in Table 5. System response time and utilization.}
	\label{fig13a}
\end{figure}

\renewcommand{\thefigure}{13b}
\begin{figure}[ht]
	\centering
	\begin{subfigure}{0.45\textwidth}
		\centering
		\includegraphics[width=\textwidth]{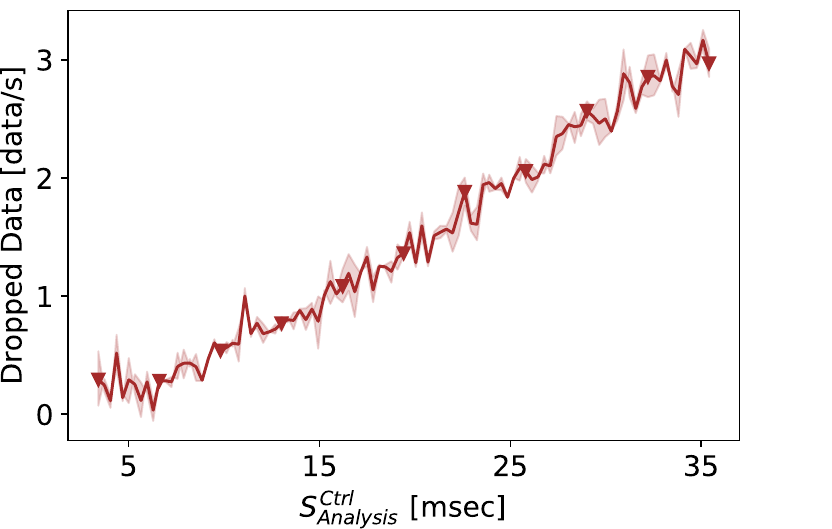}
		\caption{Original.}
	\end{subfigure}
	\hfill
	\begin{subfigure}{0.45\textwidth}
		\centering
	    \includegraphics[width=\textwidth]{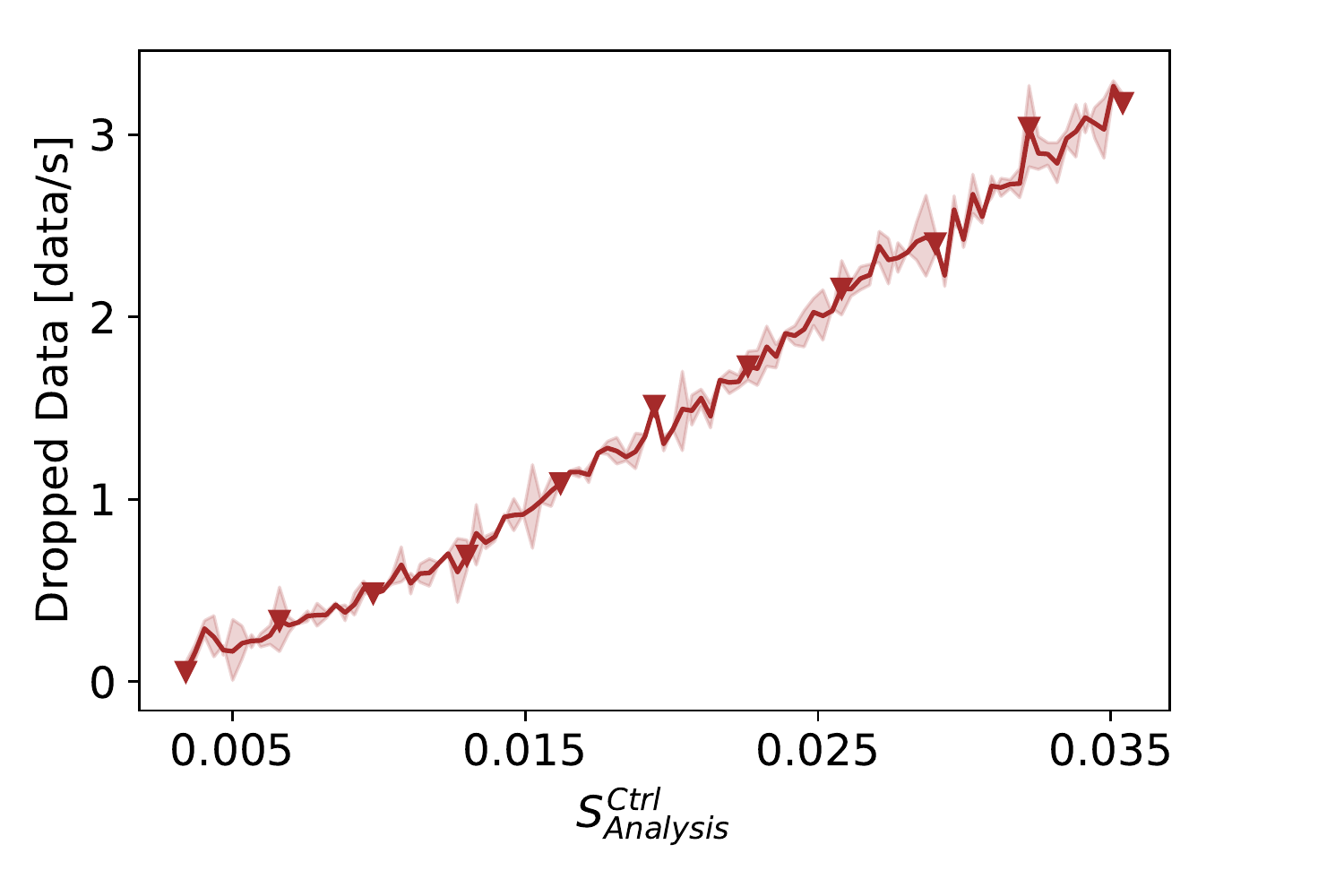}
		\caption{Reproduced.}
	\end{subfigure}
	\caption{Effect of \textit{Where Was I?} software performance antipattern on the sensor net system. Results are obtained considering only 1 Actor and 1 Sensor in the system. Other input parameters are the same as those in Table 5. Dropped data.}
	\label{fig13b}
\end{figure}

\renewcommand{\thetable}{6}
\begin{table}[!htb]
    \caption{Execution graph (EG) and queueing network (QN) results to assess the correctness of the adopted models used to study the CPS. The 99\% confidence interval of JMT simulations is shown in parenthesis. The utilization error is the distance between the observed usages. The system response time error is a mean absolute percentage error. Errors are computed wrt. average values.}
    \label{thetab}
    \centering
      Original:\vskip10pt
        \begin{tabular}{c||ccc|ccc}
            & \multicolumn{3}{c|}{Utilization} & \multicolumn{3}{c}{System Response Time} \\
            \hline
            Job Class & EG [\%] & QN [\%] & Error [\%] & EG [msec] & QN [msec] & Error [\%] \\
            \hline
            Analysis & 17.4 & 17.8 ($\pm 0.41$) & 0.4 & 5.53 & 5.35 ($\pm 0.10$) & 3.18 \\
            Status & 3.9 & 4.1 ($\pm 0.08$) & 0.2 & 1.17 & 1.11 ($\pm 0.02$) & 5.05 \\
            Actors & 16.1 & 15.8 ($\pm 0.46$) & 0.3 & 3.51 & 3.64 ($\pm 0.07$) & 3.85 \\
            Polling & 10.0 & 10.9 ($\pm 0.30$) & 0.9 & 2.06 & 2.18 ($\pm 0.04$) & 5.72 \\ \hline
        \end{tabular}
      \vskip30pt
      Reproduced:\vskip10pt
        \begin{tabular}{c||ccc|ccc}
            & \multicolumn{3}{c|}{Utilization} & \multicolumn{3}{c}{System Response Time} \\
            \hline
            Job Class & EG [\%] & QN [\%] & Error [\%] & EG [msec] & QN [msec] & Error [\%] \\
            \hline
            Analysis & 17.40 & 18.18 ($\pm 0.42$) & 0.78 & 5.38 & 5.33 ($\pm 0.10$) & 0.92 \\
            Status & 4.00 & 4.10 ($\pm 0.08$) & 0.10 & 1.10 & 1.12 ($\pm 0.02$) & 1.78 \\
            Actors & 16.10 & 15.95 ($\pm 0.45$) & 0.15 & 3.47 & 3.62 ($\pm 0.07$) & 4.14 \\
            Polling & 11.00 & 10.80 ($\pm 0.27$) & 0.20 & 2.10 & 2.13 ($\pm 0.06$) & 1.41 \\ \hline
        \end{tabular}
\end{table}



\begin{thebibliography}{3} 

\bibitem{paper} Riccardo Pinciroli, Connie U. Smith, and Catia Trubiani, ``QN-based Modeling and Analysis of Software Performance Antipatterns for Cyber-Physical Systems," in \emph{Proceedings of the 12$^{\text{th}}$ ACM/SPEC International Conference on Performance Engineering} (ICPE). ACM, 2021.

\bibitem{JMT} Marco Bertoli, Giuliano Casale, and Giuseppe Serazzi. ``Java modelling tools: an open source suite for queueing network modelling andworkload analysis," in \emph{Proceedings of the Third International Conference on the Quantitative Evaluation of Systems} (QEST). IEEE, 2006.

\bibitem{SPEED}
Connie U. Smith, and Lloyd G. Williams. ``Performance engineering evaluation of object-oriented systems with SPE$\cdot$ED\texttrademark," in \emph{Proceedings of the 1997 International Conference on Modelling Techniques and Tools for Computer Performance Evaluation}. Springer, Berlin, Heidelberg, 1997.

\end{thebibliography}
\end{document}